# Title: P2P Networks for Content Sharing

## *Authors:*

Choon Hoong Ding, Sarana Nutanong, and Rajkumar Buyya

Grid Computing and Distributed Systems Laboratory,
Department of Computer Science and Software Engineering,
The University of Melbourne, Australia
(chd, sarana, raj)@cs.mu.oz.au

## *ABSTRACT*

Peer-to-peer (P2P) technologies have been widely used for content sharing, popularly called "file-swapping" networks. This chapter gives a broad overview of content sharing P2P technologies. It starts with the fundamental concept of P2P computing followed by the analysis of network topologies used in peer-to-peer systems. Next, three milestone peer-to-peer technologies: Napster, Gnutella, and Fasttrack are explored in details, and they are finally concluded with the comparison table in the last section.

# 1. INTRODUCTION

Peer-to-peer (P2P) content sharing has been an astonishingly successful P2P application on the Internet. P2P has gained tremendous public attention from Napster, the system supporting music sharing on the Web. It is a new emerging, interesting research technology and a promising product base.

Intel P2P working group gave the definition of P2P as "The sharing of computer resources and services by direct exchange between systems". This thus gives P2P systems two main key characteristics:
- Scalability: there is no algorithmic, or technical limitation of the size of the system, e.g. the complexity of the system should be somewhat constant regardless of number of nodes in the system.
- Reliability: The malfunction on any given node will not effect the whole system (or maybe even any other nodes).

File sharing network like Gnutella is a good example of scalability and reliability. In Gnutella, peers are first connected to a flat overlay network, in which every peer is equal. Peers are connected directly with out need of a control master server's arrangement. And the malfunction of any node does not cause any other nodes in the system any trouble.

P2P can be categorized into two groups classified by the type of model: pure P2P, and hybrid P2P. Pure P2P model, such as Gnutella, Freenet, does not have a central server. Hybrid P2P model, such as Napster, Groove, Magi, employs a central server to obtain meta-information such as the identity of the peer on which the information is stored or to verify security credentials. In a hybrid model, peers always contact a central server before they directly contact other peers.

## 2. P2P Networks Topologies

According to (Peter, 2002), all peer-to-peer topologies, no matter how different they may be, will have one common feature. All file transfers made between peers are always done directly through a data connection that is made between the peer sharing the file and the peer requesting for it. The control process prior to the file transfer, however, can be implemented in many other ways. As stated by (Nelson, 2001), P2P file sharing networks can be classified into four basic categories which are the centralized, decentralized, hierarchical and ring systems. Although these topologies can exist on their own, it is usually the practice for distributed systems to have a more complex topology by combining several basic systems to create, what is known now as hybrid systems. We will give a brief introduction to the four basic systems and later delve deeper into the topic of hybrid systems.

## *2.1 Centralized Topology*

The concept of a centralized topology is very much based on the traditional client/server model. Please refer to Figure 1 for illustration. A centralized server must exist which is used to manage the files and user databases of multiple peers that log onto it (Peter, 2002). The client contacts the server to inform it of its current IP address and names of all the files that it is willing to share. This is done every time the application is launched. The information collected from the peers will then be used by the server to create a centralized database dynamically, that maps file names to sets of IP addresses.

All search queries will be sent to the server, who will perform a search through its locally maintained database. If there is a match, a direct link to the peer sharing the file is established and the transfer executed (Kurose, 2003). It should be noted here that under no circumstances, will the file ever be placed on the server.
Examples of applications that make use of such a network would be seti@home, folding@home, Napster which will be discussed in greater detail in the following sections, and many more.

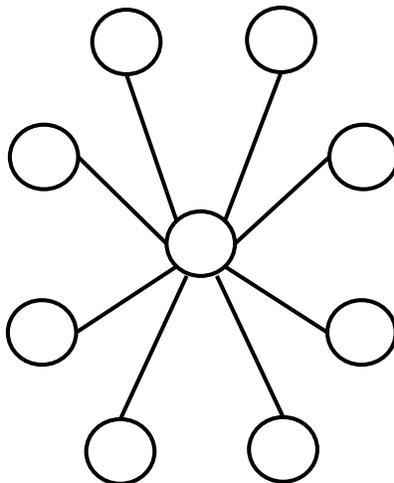
Figure 1: Illustration of the Centralized Topology

## *2.2 Ring Topology*

It should be relatively clear that the drawback of a centralized topology is that the central server can become a bottle neck (when load becomes heavy) and a single point of failure. These are some of the main contributing factors as to why the ring topology came about.

It is made up of a cluster of machines that are arranged in the form of a ring to act as a distributed server (Nelson, 2001). This cluster of machines will work together to provide better load balancing and high availability. This topology is generally used when all the machines are relatively nearby on the network, which means that it is most likely owned by a single organization; where anonymity is not an issue. Figure 2 shows a simple illustration of a ring topology.

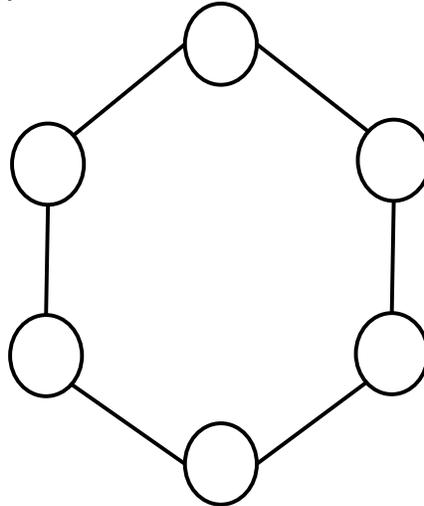
Figure 2: Illustration of the Ring Topology

## *2.3 Hierarchical Topology*

Hierarchical systems have been in existence for a very long time. Even on the Internet, hierarchical systems have been about since the very beginning. Many Internet applications function in a hierarchical environment. The best example of a hierarchical system on the Internet would be the Domain Name Service (DNS) (Nelson, 2001). Authority flows from the root name servers to the servers of the registered name and so on so forth. This sort of topology is very suitable for systems that require a form of governance that involves delegation of rights or authority. Another good example of a system that makes use of the hierarchical topology would be the Certification Authorities (CA) that certify the validity of an entity on the Internet. The root CA can actually delegate some of its authoritative rights to companies that subscribe to it, so that those companies can, in turn grant certificates to those that reside underneath it. Figure 3 provides a brief illustration of how a hierarchical system looks like.

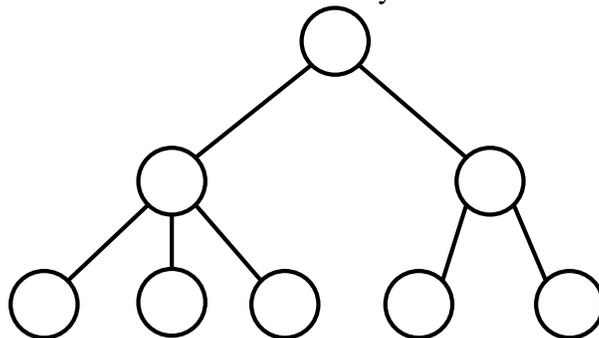
Figure 3: Illustration of the Hierarchy Topology

## *2.4 Decentralized Topology*

In a pure peer-to-peer architecture, no centralized servers exist. All peers are equal, hence creating a flat, unstructured network topology (Peter, 2002). Please refer to Figure 4 for illustration. In order to join the network, a peer must first, contact a bootstrapping node (node that is always online), which gives the joining peer the IP address of one or more existing peers, officially making it part of the ever dynamic network. Each peer, however, will only have information about its neighbors, which are peers that have a direct edge to it in the network.

Since there are no servers to manage searches, queries for files are flooded through the network (Kurose, 2003). The act of query flooding is not exactly the best solution as it entails a large overhead traffic in the network.

A good example of an application that uses this model is Gnutella. Details of how it searches and shares files in a pure peer-to-peer network will be discussed later.

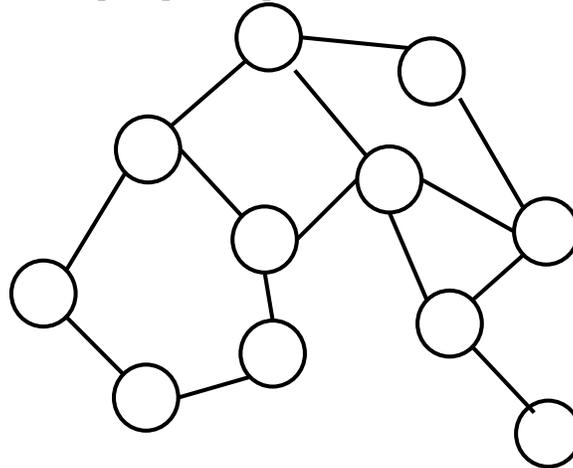

Figure 4: Illustration of the Decentralized Topology

## *2.5 Hybrid Topology*

Having discussed the basic topologies of peer-to-peer networks, we now come to the more complex real world systems that generally combine several basic topologies into one system. This is known as the Hybrid architecture (Yang, 2002). We will discuss several of such examples in this section just to give a brief idea of this sort of architecture. In such a system, nodes will usually play more than one role.

### **2.5.1 Centralized Topology and Ring Topology**

This hybrid topology is a very common sight in the world of web hosting (Nelson, 2001). As mentioned previously in the ring topology section, heavy loaded web servers usually have a ring of servers that specializes in load balancing and failover. So, the servers themselves maintain a ring topology. The clients however are connected to the ring of servers through a centralized topology (i.e. client/server system). Therefore, the entire system is actually a hybrid; mixture between the sturdiness of a ring topology with the simplicity of a centralized system. Figure 5 gives a simple illustration of such a topology.

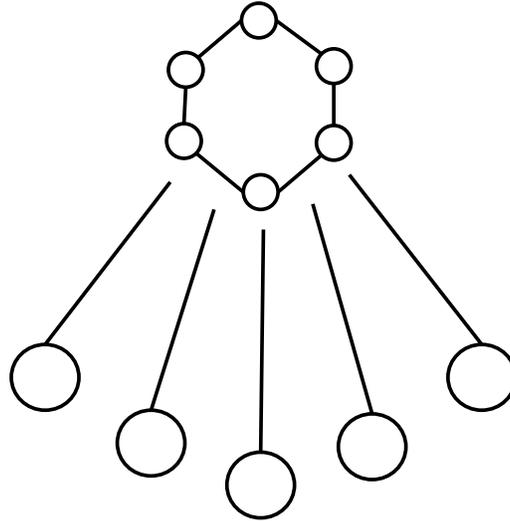

Figure 5: Illustration of the Centralized and Ring Topology

## 2.5.2 Centralized Topology and Centralized Topology

It is often the case where the server of a network is itself a client of a bigger network (Nelson, 2001). This sort of hybrid topology is a very common practice in organizations that provide web services. A simple example that will help illustrate this point would be, when a web browser contacts a centralized web server (Refer to Figure 6). The web server may process and format the results so that they can be presented in HTML format and in the process of doing that, these servers might themselves contact other servers (e.g. Database server) in order to obtain the necessary information (Nelson, 2002).

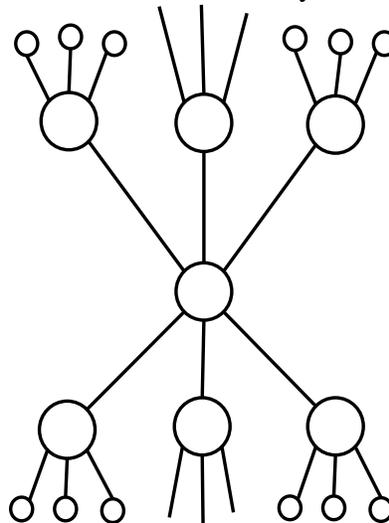

Figure 6: Illustration of the Centralized and Centralized Topology

## 2.5.3 Centralized Topology and Decentralized Topology

In this topology, peers that function as group leaders are introduced (Kurose, 2003). They have been known by many names. Some call them Group Leader Nodes, Super Nodes or even Ultra Nodes. To keep things simple and consistent with the following sections about Kazaa, we will refer to them as Super Nodes from here onwards.

These Super Nodes will perform the task of a centralized server as in the centralized topology, but only for a subset of peers. The Super Nodes themselves are tied together in a decentralized manner. Therefore, this hybrid topology actually introduces two different tiers of control. The first is where ordinary peers connect to the Super Nodes in a centralized topology fashion. The second is where the Super Nodes connect to each other in a decentralized topology fashion. Please refer to Figure 7.

As with the centralized topology, the Super Nodes maintains a database that maps file names to IP addresses of all peers that are assigned to it (Yang, 2002). It should be noted here that the Super Node's database only keeps track of the peers within its own group. This greatly reduces the scope of peers that it needs to serve. So, any ordinary peer with a high speed connection will qualify to be a Super Node. The best example of a peer-to-peer application that utilizes such a topology would be Kazaa/FastTrack.

Another good example of such a topology would be the common Internet email. Mail clients have a decentralized relationship to specific mail servers. Like the Super Nodes, these mail servers share emails in a decentralized fashion among themselves.

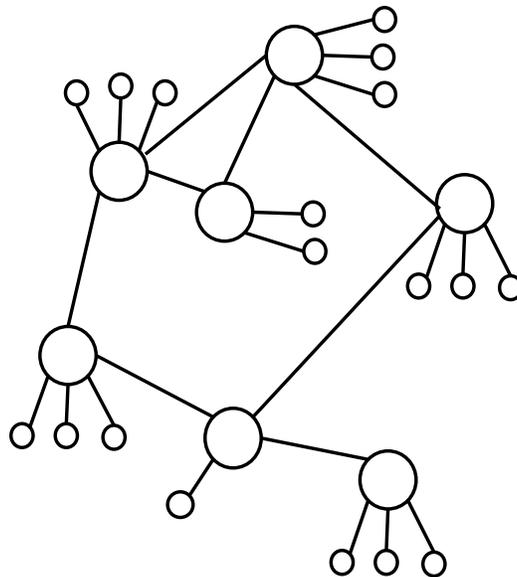

Figure 7: Illustration of the Centralized and Decentralized Topology

### 2.5.4 Other Potential Topologies

Please take note that the hybrid topologies mentioned so far are just the common ones. As can be seen, there can be a great deal of different combinations of hybrid topologies that can be achieved from the basic topologies. However, if one were to make too many combinations the resulting topology may become too complex hence making it difficult to manage.

# 3. Napster

Napster is a file-sharing P2P application that allows people to search for and share MP3 music files through the vast Internet. It was single handedly written by a teenager named Shawn Fanning (Jeff, 2000). Not only did he develop the application, but he also pioneered the design of a protocol that would allow peer computers to communicate directly with each other. This paved a way for more efficient and complex P2P protocols by other organizations and groups.

## *3.1 The Napster Architecture*

The architecture of Napster is based on the Centralized Model of P2P file-sharing (Martin, 2000). It has a Server-Client structure where there is a central server system which directs traffic between individual registered users. The central servers maintain directories of the shared files stored on the respective PCs of registered users of the network. These directories are updated every time a user logs on or off the Napster server network. Clients connect automatically to an internally designated "metaserver" that acts as common connection arbiter. This metaserver assigns at random an available, lightly loaded server from one of the clusters. Servers appeared to be clustered about five to a geographical site and Internet feed, and able to handle up to 15,000 users each. The client then registers with the assigned server, providing identity and shared file information for the server's local database. In turn, the client receives information about connected users and available files from the server. Although formally organized around a user directory design, the Napster implementation is very data centric. The primary directory of users connected to a particular server is only used indirectly, to create file lists of content reported as shared by each node (David, 2001).

Users are almost always anonymous to each other; the user directory is never queried directly. The only interest is to search for content and determine a node from which to download. The directory therefore merely serves as a background translation service, from the host identity associated with particular content, to the currently registered IP address needed for a download connection to this client. Each time a user of a centralized P2P file sharing system submits a request or search for a particular file, the central server creates a list of files matching the search request, by cross-checking the request with the server's database of files belonging to users who are currently connected to the network. The central server then displays that list to the requesting user. The requesting user can then select the desired file from the list and open a direct HTTP link with the individual computer which currently posses that file. The download of the actual file takes place directly, from one network user to the other, without the intervention of the central server. The actual MP3 file is never stored on the central server or on any intermediate point on the network.

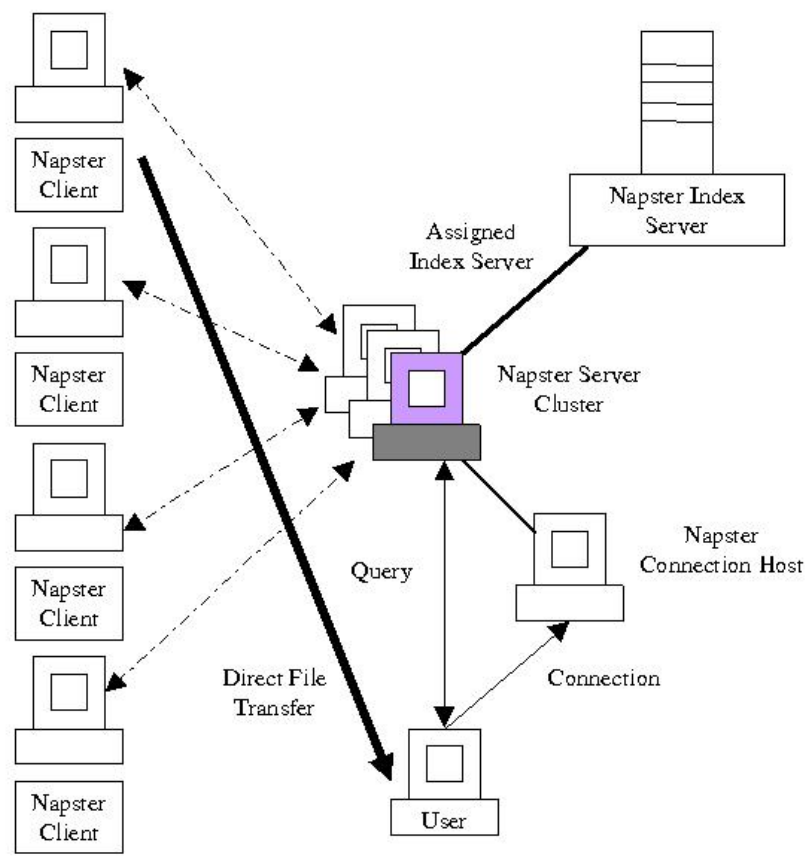

Figure 8: Illustration of the Napster Architecture

## 3.2 The Napster Protocol

Due to the fact that Napster is not an open source application, it was only possible to build up a similar application in revealing the Napster protocol by reverse-engineering (Eduard, 2002). In other words, no one will ever be totally sure how the Napster protocol specification is like, except for the creator of Napster himself. Project OpenNap has made it possible to run a Napster server on many platforms without using the original Napster application and the index server. The following are the protocol specification for Napster with reference to (Eduard, 2002).

Napster works with a central server which maintains an index of all the mp3 files of the peers. To get a file you have to send a query to this server which sends you the port and IP address of a client sharing the requested file. With the Napster application it is now possible to establish a direct connection with the host and to download a file.

In contrast to Gnutella who only uses a handful of messages to communicate between peers, the Napster protocol uses a whole lot of different types of messages. Every state of the hosts, acting like clients towards the server, is related to the central Napster server. Thus the Napster protocol makes anonymity impossible. At first, one may think that this is a drawback, but this complex protocol actually makes a lot of services possible. Some examples are:
- Creating Hotlists: notifying when users of your own hotlist sign on or off the server
- List of ignored User
- Instant Messaging: sending public or private messages to other users; creating and joining channels of shared interests

### 3.2.1 Napster Messages Data Structures

Each message to/from the Napster central server is of the form in the figure below:

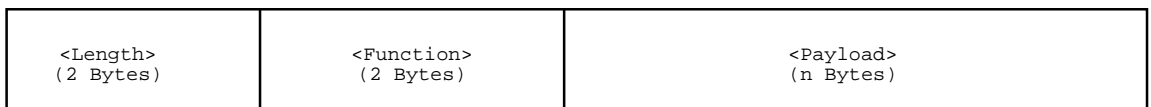

Figure 9: Messages Data Structures

Where:
- Length specifies the length of the payload.
- Function defines the message type of the packet (see next paragraph)
- Payload this portion of the message is a plain ASCII string

Every block of header and payload is separated by "blanks" which make the synchronization of the incoming messages possible. Most of blocks have no fixed length. The blanks make separation of data blocks in incoming bit streams possible.

### 3.2.2 Initialisation

A registered Napster host, acting like a client, sends to the server a LOGIN(0x02) message with the following format:

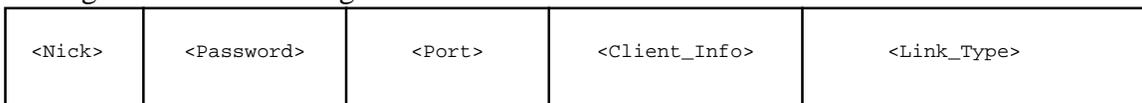

Figure 10: Login Message Structure

Where:
- Nick & Password identify the user
- Port is the port which the client is listening on for data transfer.
- Client_Info is a string containing the client version info.
- Link_Type}is a integer indicating the client's bandwidth.

The details are given in Table 1:

| Representation | Bandwidth | Representation | Bandwidth |
| --- | --- | --- | --- |
| 0 | unknown | 1 | 14.4 kbps |
| 2 | 28.8 kbps | 3 | 33.6 kbps |
| 4 | 56.7 kbps | 5 | 64k ISDN |
| 6 | 128k ISDN | 7 | Cable |
| 8 | DSL | 9 | T1 |
| 10 | T3 or greater | | |

Table 1: Assessment Details

The host's IP address hasn't to be added to the message. However, the server can extract it automatically from the TCP packet in which the message is packed for the transmission.

An unregistered host sends a New User Login (0x06) which is similar to the format of Login (0x02), with the addition of the email address on the end. The server sends a Login Ack (0x03) to the client after a successful login. If the nick is registered, the email address given at registration time is returned, else, a dummy value will be returned.

### 3.2.3 Client Notification of Shared File (0x64)

With the Client Notification of Shared File (0x64) message the client sends successively all the files it wants to share.

| `<Filename>` | `<MD5>` | `<Size>` | `<Bitrate>` | `<Frequency>` | `<Time>` |
| --- | --- | --- | --- | --- | --- |

Figure 11: Client Notification of Shared File (0x64)

- MD5 is the hash value of the shared file. The MD5 (Message Digest 5) algorithm produce a 128-bit "fingerprint" of any file. It is nearly computationally infeasible to produce two messages having the same hash value. The MD5 algorithm is intended to provide any user the possibility to secure the origin of his shared file, even if the file is laying on drives of other Napster users.
- Size is the file size in bytes
- Bitrate is the bit rate of the mp3 in kbps
- Frequency is the sample rate of the mp3 in Hz
- Time is the duration of the music file in seconds

### 3.2.4 File Request

The downloading client will first issue either a Search (0xC8) or Browse(0xD3). The first message has the following format:

| `<Artist Name>` | `<Title>` | `Bit-rate` | `<Max Results>` | `<Line-type>` | `<Frequency>` |
| --- | --- | --- | --- | --- | --- |

Figure 12: Search Message (0xC8)

- Max is the maximum number of results.
- Link-Type Range is the rage of link-types.
- Bit-rate Range is the rage of bit-rate.
- Frequency Range is the range of sample frequencies in Hz.

The artist name and the song title are checked from the file name only. Napster does not make use of the ID3 in mp3 files in its search criteria.
The payload of the Browse (0xD3) message does only contains the <nick> of the host. It requests a list of the host's shared files.

### 3.2.5 Response and Browse Response

The server answers respectively with a Search Response (0xC9) or a Browse Response (0xd4) with the formats given in the figure below:

| <Filename> | <MD5> | <Size> | <Bit-rate> | <Frequency> | <Time> | <Nick> | <IP> | <Link-type> |
|---|---|---|---|---|---|---|---|---|

Figure 13: Response Message

Where:
- MD5 hash value of the requested file
- Size file size in bytes
- Bitrate bit rate of the mp3 in kbps
- Frequency sample rate of the mp3 in Hz
- Time specify the length of the file
- Nick identify the user who shares the file
- IP 4 Bytes integer representing the IP address of the user with the file.
- Link-Type Refer to Login Message.

### 3.2.6 Download Request

To request a download, a DOWNLOAD REQUEST (0xCB) message is sent to the server. The client requests to download <filename> from <nick>. This message has the following payload format:

| <Nick> | <Filename> |
|---|---|

Figure 14: Download Request Message

### 3.2.7 Download ACK

The server will answer with a DOWNLOAD ACK (0xCC) containing more information about the file (Linespeed, Port Number, etc). This message has the following payload format:

| <Nick> | <IP> | <Port> | <Filename> | <MD5> | <Link-type> |

Figure 15: Download Acknowledgement Message

### 3.2.8 Alternate Download Request

It is like the normal "Download Request", only difference is that it is for use when he person sharing the file can only make outgoing TCP connection because of the firewall that is blocking the incoming messages. The ALTERNATE DOWNLOAD REQUEST (0x1F4) message should be used to request files from users who have specified their data port as '0' in their login message.

### 3.2.9 Alternate Download Ack

This ALTERNATE DOWNLOAD ACK (0x1F5) is sent to the uploader when its data port is set to 0 to indicate they are behind a firewall and need to push all data. The uploader is responsible for connecting to the downloader t transfer the file.

### 3.2.10 File Transfer

From this point onwards, the hosts don't send messages to the central server anymore. The host requesting the file makes a TCP connection to the data port specified in the 0xCC message from the server. To request for the file that the client wish to download, it sends the following HTTP - messages: a string "GET" in a single packet and a message with the format:

| <Nick> | <Filename> | <Offset> |

Figure 16: Request Message

- Nick is the client's nick.
- Offset is the byte offset in the file to begin the transfer at. It is needed to resume prior transfer.

The remote host will then return the file size and, immediately following, the data stream. The direct file transfer between Napster hosts uses a P2P architecture. Once the data Transfer is initiated, the downloader should notify the server that they are downloading a file by sending the DOWNLOADING FILE (0xDA) message. Once the transfer is complete, the client send a DOWNLOAD COMPLETE (0xDB) message.

### 3.2.11 Firewalled Downloading

Napster also has method to allow clients behind firewalls to share their contents as well. As described above, when the file needs to be pushed from a client behind a firewall, the downloader sends a message ALTERNATE DOWNLOAD REQUEST (0x1F4) message to the server. This causes an ALTERNATE DOWNLOAD ACK (0x1F5) to be sent to the uploader, which is similar to the DOWNLOAD REQUEST (0xCB) message for a normal download.

Once the uploader receives the (0x1F5) message from the server, it should make a TCP connection to the downloader's (0x1F5) data port (given in the message). Upon connection, the downloader's client will sent one byte, the ASCII character '1'. The uploader should then send the string "SEND" in a single packet, and then the message (format was shown before).

Upon receipt, the downloading client will either send the byte offset at which the transfer should start, or an error message such as "INVALID REQUEST". The byte offset should be sent as a single packet in plain ASCII digits. A 0 byte offset indicates the transfer should begin at the start of the file.

## *3.3 Implementation*

The Napster protocol was a closed one, meaning no one knows for sure how file searching and transfer is done. So, when Napster was first introduced, there was only one client implementation, which was called the Napster, for obvious reasons.

Napster exclusively focuses on MP3-encoded music files. Although no other file types were supported, an intriguing subculture of client clones and tools soon arose, reverse-engineered from the Napster's closed-source clients.

The intent behind the development was to have greater user control. For instance, a form of MP3 spoofing implemented by tools such as Wrapster could enclose an arbitrary file with a kind of wrapper that made it look like an MP3 file to the Napster servers. It would then appear in the server databases and be searchable by other clients wishing to download files other than music. An obstacle to this effort was that the artist-title description field allowed little information about the non-music file. This, the low ratio of non-music to music files, and the normal random distribution of connecting nodes conspired to make Napster's scope-limited searches highly unlikely to find special content. Tools such as Napigator were developed to allow users to connect to specific servers, bypassing metaserver arbitration. In this way, certain servers became known as Wrapster hangouts-primary sites for non-music content. Users looking for this kind of content were then more likely find it.

Nonmusic exchanges over Napster proper were never more than marginal, at least compared to alternative, content-agnostic systems such as Gnutella. Some alternative Napster servers such as OpenNap started as "safe-havens" for Napster users when Napster began filtering content, did for a while begin to fill the gap, tying together former

Napster clients, clones and variations with a new kind of server that extended the original Napster protocol to all file types. No matter how much the Napster model was reengineered, however, the fundamental requirement of a "Napster-compatible" central server remained a serious constraint for a network based on this technology or any of its clones. To transcend this limitation, other protocols and architecture models are needed- for example, serverless networks in the style of Gnutella.

## 4. Gnutella

In the early of March 2000, Gnutella was originated by Justin Frankel and Tom Pepper, working under the Gnullsoft, which is one of the AOL subsidiaries. However, Gnutella's development was halted by AOL shortly after it was published, but during that time several curious programmers already completed downloading it. Thanks to those downloadings, many open-source developers quickly reverse-engineered Gnutella's communication protocol and published a number of Gnutella clones with several improvements, e.g., LimeWire, BearShear, Gnucleus, XoloX, and Shareaza.

### *4.1 The Gnutella Architecture*

Instead of using a centralized index directory cluster of servers, Gnutella uses
a flat network of peers, servents, to maintain the index directory of all of the content
in the system.

In a Gnutella network, servents are connected to each other in a flat ad-hoc topology,
or a P2P networks for servents. A servent works like a client and a server by itself. As a server, it responses to queries from another peer servent. As a client, it issues queries to other peer servents.

For a servent to join the Gnutella network, it must find the address of a servent that is already connected to the network. This can be done by using host caches, such as GnuCache, which caches Gnutella servents (hosts) that always connect to the Gnutella network. After an address if found, it then sends a request message GNUTELLA CONNECT to the already connected servent. The requested servent may either accept the request by sending a reply message GNUTELLA OK, or reject the request message by sending any other response back to the requesting servent. A rejection can happen due to different reasons such as, an exhaustion of connection slots, having different versions of the protocol, etc (Limewire). Once attached to the network, the servent periodically pings its neighbors to discover other servents. Typically, each servent should connect to more than one servent since the Gnutella network is dynamic, which means any servents can go off-line or disconnect any time.

It is thus important to stay in contact with several servents at the same time to prevent being disconnected from the network. Once a server receive the ping message, it sends back a pong message to the server that originated the ping message using the same path that the ping message came from. A pong message contains the details of the servent like port, IP address, the number of files shared, and the number of kilobytes shared.

Gnutella dose not use any directory servers, as each servent maintain their local index directory. To search for a file, a query is sent to the neighbors Once a query is received by a neighbor, the query criteria is checked against the local index directory and propagate this query message to their neighbors, and so forth.
If the check matches with the local data in any servent, the servent will send back the queryHit message to the query initiating servent along the same path that carried the query message. However, when the servent generating the queryHit message stays behind the firewall, the requesting servent cannot create a connection to it. In this case, the push message will be sent by the requesting servent to the servent that generates the queryHit message and stays behind the firewall to initiate the connection instead. Note that file transfer is done by HTTP protocol, not by the Gnutella protocol.

To prevent flooding the network with the messages, the TTL (Time-To-Live) field is included in the header of every Gnutella message. The TTL field will be decremented by one every time the message is passed one servent. The servent decrementing the TTL and finding that its value equals to zero will drop that message. Each servent also needs to maintain the list of recently seen messages by storing the Descriptor ID and the Payload Descriptor of each incoming message to prevent forwarding the same message repeatedly. The detailed description of the Gnutella message and protocol will be explained in the next section.

To better illustrate the concept of Gnutella protocol, we will give one concrete example of Gnutella protocol.

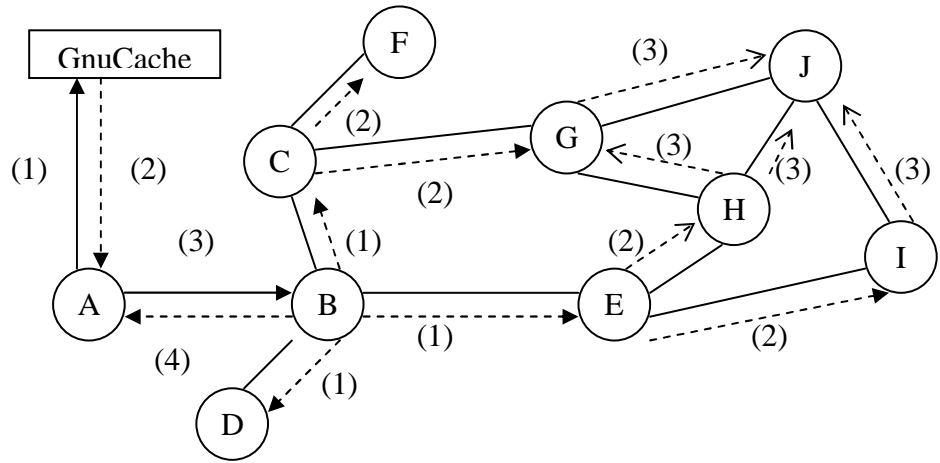

(1) User A connects to the GnuCache to get the list of available servents already connected in the network
(2) GnuCache sends back the list to the user A
(3) User A sends the request message GNUTELLA CONNECT to the user B
(4) User B replies with the GNUTELLA OK message granting user A to join the network

Figure 17: Example of Gnutella Protocol

From the figure above, suppose that once user A connected to the network, he wants to search for some files. Therefore, he sends a query message to his neighbor, user B. User B first checks that this is not the old message, then checks for matching with his local data. If matches, he sends the queryHit message back to user A. User B decrements TTL by 1 and forwards the query message to user C, D, and E. User C, D, and E do the same things as user B and forward the query message further to user F, G, H, and I. User F, G, H, and I also do the same things, but suppose that user H is the first person who forwards the query message to user J. The subsequent query messages forwarded by user G and I to user J will be discarded by user J when he checks against his local table and finds that he already received this message. This is hold for user G as well when user H forwards the query message to him. Suppose now that user J finds matching against his local data, he sends the queryHit message back to user A by following the same path as the query message carried, that is from J, to H, to E, to B, and to A. Now, user A can initiate the file down load directly with user J by using HTTP protocol.

## 4.2 The Gnutella Protocol

This section describes the detail of messages and the rules used in Gnutella protocol with reference to (Limewire, Ian). The message used to communicate between the servents is called Gnutella descriptors. Gnutella descriptors consist of Descriptor Header and Descriptor Payload. There are five types of Gnutella Descriptors: Ping, Pong, Query, queryHit, and Push as discussed in Table 2.

| Descriptor | Description |
|---|---|
| Ping | Used to actively discover hosts on the network. A servent receiving a Ping descriptor is expected to respond with one or more Pong descriptors. |
| Pong | The response to a Ping. Includes the address of a connected Gnutella servent and information regarding the amount of data it is making available to the network. |
| Query | The primary mechanism for searching the distributed network. A servent receiving a Query descriptor will respond with a QueryHit if a match is found against its local data set. |
| QueryHit | The response to a Query. This descriptor provides the recipient with enough information to acquire the data matching the corresponding Query. |
| Push | A mechanism that allows a firewalled servent to contribute file-based data to the network. |

Table 2: Five descriptors used in Gnutella protocol

### 4.2.1 Descriptor Header

Descriptor Header consists of five parts: Descriptor ID, Payload Descriptor, TTL, Hops, and Payload Length.

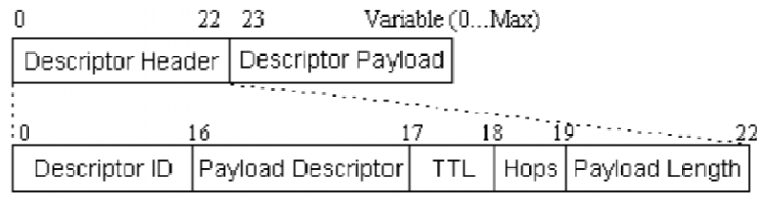

Figure 18: Descriptor Header

- Descriptor ID is a unique identifier for the descriptor on the network.
- Payload Descriptor identifies the type of each descriptor: 0x00 for Ping, 0x01 for Pong, 0x40 for Push, 0x80 for Query, and 0x81 for QueryHit.
- TTL represents the number of times the descriptor will be forwarded by Gnutella servents before it is discarded from the network.
- Hops represent the number of times the descriptors has been forwarded. The result of the current hops value plus the current TTL value always equals to the initial TTL value.
- Payload Length is used to locate the next descriptor. The next descriptor header is located exactly Payload_Length bytes from the end of this header.

### 4.2.2 Descriptor Payload

There are five types of descriptor payload: Ping, Pong, Query, QueryHit, and Push.
  1. Ping (0x00)

Ping descriptor has no associated payload and is of zero length. A servent uses Ping descriptors to actively probe the network for other servents.

  2. Pong (0x01)

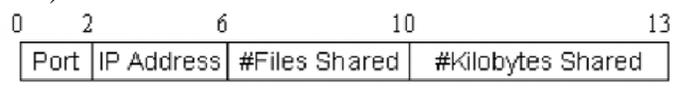

Figure 19: Pong (0x01)

Pong descriptor payload has four parts: port, IP address, the number of files shared, and the number of kilobytes shared. Pong descriptors are only sent in response to an incoming Ping descriptor. One Ping descriptor can be replied with many Pong descriptors.
Port is the port number on which the responding host can accept incoming connections.
- IP Address is the IP address of the responding host.
- #Files Shared is the number of files that the servent with the given IP address and port is sharing on the network.

- #Kilobytes Shared is the number of kilobytes of data that the servent with the given IP address and port is sharing on the network.

3. Query (0x80)

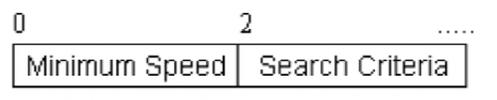

Figure 20: Query (0x80)

Query descriptor, the Gnutella's search message format, consists of two parts: minimum speed, and search criteria. Minimum Speed is the minimum speed in KBytes/sec of the servent that should respond to this message. Search Criteria contains the search criteria of the requesting servent. The maximum length of the search criteria is bounded by the Payload_Length field of the descriptorheader.

4. QueryHit (0x81)

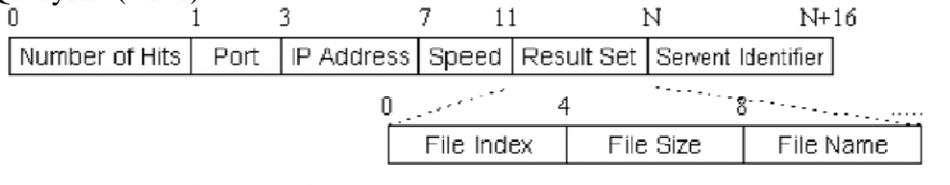

Figure 21: QueryHit (0x81)

QueryHit descriptors are the responses to the Query descriptor by the servents when they find matching against their local data. The descriptor ID field in the Descriptor Header of the QueryHit should be the same as that of the associated Query descriptor. This will allow the requesting servent to identify the QueryHit descriptor associated with the Query descriptor it generated.

- Number of Hits is the number of query hits in the result set.
- Port is the port number on which the responding host can accept incoming connections.
- IP Address is the IP address of the responding host.
- Speed is the speed in KBytes/sec of the responding host.
- Result Set is a set of responses to the corresponding Query. This set contains the Number_of_Hits elements, each of which has the structure comprising file index, file size, and file name.
- Servent Identifier is a 16-byte string uniquely identifying the responding servent on the network.

5. Push (0x40)

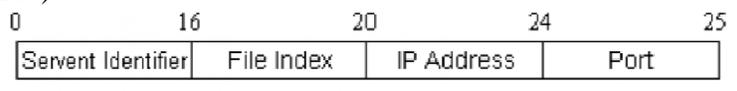

Figure 22: Push (0x40)

Push descriptor is used by the requesting servent to ask for a responding servent behind the firewall to initiate the connection.

- Servent Identifier is a 16-byte string uniquely identifying the servent who is being requested to push the file with index File_Index. This servent identifier is set to the same as the servent identifier returned in the corresponding QueryHit descriptor sending by the servent behind the firewall.
- File Index uniquely identifies the file to be pushed from the requested servent.
- IP Address is the IP address of the requesting host.
- Port is the port number of the requesting host.

### 4.2.3 Rules

There are generally 5 rules for Gnutella protocol mandating the servents to strictly follow to maintain desirable network traffics.

- Rule 1: All Servents must memorize the unique 128-bit Descriptor ID every time a message is delivered or originated. If these memorized messages are received again, it will not be forwarded. This helps eliminating looping in the network thereby reducing the unnecessary traffic.
- Rule 2: Pong, QueryHit, and Push descriptors may only be sent along the same path that carried the incoming Ping, Query, and QueryHit descriptors, respectively. This ensures that only those servents that routed the Ping (Query, and QueryHit) descriptor will see the Pong (QueryHit, and Push) descriptor in response. A servent that receives a Pong (QueryHit, and Push) descriptor with Descriptor ID = n, but has not seen a Ping (Query, and QueryHit) descriptor with Descriptor ID = n should discard the Pong (QueryHit, and Push) descriptor from the network.
- Rule 3: A servent will forward incoming Ping and Query descriptor to all of its directly connected servents, except for the one that sent the incoming Ping or Query.
- Rule 4: A servent will decrement a descriptor header's TTL field, and increment the Hops field, before it forwards the descriptor to any directly connected servent. If after decrementing the header's TTL field, the TTL field is found to be zero, the descriptor is discarded.
- Rule 5: If a servent receives a descriptor with the same Payload Descriptor and Descriptor ID as the one it has received before (check by comparing with the ones the servent stores in the table), a servent should discard this descriptor.

### *4.3 Gnutella protocol analysis and improvement methods*

As with other P2P file-sharing protocol, Gnutella was intentionally designed to achieve the four basic goals:

- Flexibility: The fact that the Gnutella network is ad-hoc, any servents can join or leave the network at anytime they want and this typically occurs frequently. Therefore, the Gnutella protocol must be designed to be flexible enough to keep operating efficiently despite the constantly changing set of servents.
- Performance and Scalability: Performance is considered in term of the throughput that any servents initiating query messages will get the QueryHit messages replying back in an acceptable time. Scalability implies that the Gnutella protocol

> should be able to handle a large number of servents without so much degradation in performance.
- Reliability: Reliability focuses on security issues that external attacks should not be able to degrade significant data or performance loss.
- Anonymity: Anonymity concerns with protecting the privacy of participants, the identity of the people seeking or providing the information.

### 4.3.1 Flexibility

In term of the dynamic environment issue, since a servent periodically pings other servents, this will prevent him from being cut from the network and will keep the network to stay functioning as long as there is a connection between any servents.
However, the strength of the Gnutella comes from the fact that there are a huge amount of users being on-line concurrently and most of them share the information.

### 4.3.2 Performance and Scalability

The performance and scalability is the big issue debated in Gnutella communities. The first problem relating to the performance issue arises from many servents connecting via the low-speed modem. These servents usually stay scattered all over the network and typically get over-flooded with the messages until they become unresponsive acting as they stay off-line. This may cause the network to be highly fragmented into isolated clusters of peers; thus, query messages will not go beyond each fragment causing the limit of search.

The second problem concerns with the ping and pong messages. Since every servent periodically pings other servents to get their addresses and there are usually a large number of servents, this indicates that the number of ping and pong messages are enormous. As investigated by Dimitri (2002), the number of pong messages consume up to about 50% of all Gnutella traffic. This will prevent other more important messages such as query, QueryHit, and push to route through the network. The consequence is that the performance of the network will eventually be degraded because most of the network bandwidths are used to send ping and pong messages (note that ping messages are boadcasted to every directly connected peers).

The third performance problem stems from the fact that Gnutella does not enforce users to share files, and the current fact is that most users do not share their files (they are free-loaders). Suppose that the TTL of the query messages is 7 and every servent is connected to four other servents. Theoretically, the search will go through $4^7 = 16,384$ servents, but if only 10% of those servents share their files, the search will drop to only around 1,600 servents. This reduces the probability to find the desired files dramatically, and this still does not consider that fact that some servents may go off-line or some servents may drop the query messages due to their congestion problem.

The fourth problem with performance is download failures which cause from the sharing of partial files or the limit of upload speed set by the people who share the files. Back to the scalability issue, (Jordan, 2001) claimed that the Gnutella network is not scalable. Due to the propagating nature of the Gnutella protocol that makes Gnutella traffics consume a lot of bandwidths, the number of Gnutella users may be limited by the underlying network bandwidth (traffic jam because of the large number of message transmitted over the network) even though Gnutella was initially designed to be used by unlimited number of users. TTL field in the Descriptor Header, the cache of previously seen message kept by each servent, and the rules imposed by the Gnutella protocol are designed to help reduce this problem, but more powerful techniques are needed to ensure true scalability.

There are several approaches (discussed based on (LimeWire, 2003; Igor) proposed to help achieve higher performance and scalability: encouragement of content sharing, blocking freeloaders, reducing unnecessary network traffic, creating and maintaining a healthy network structure. Encouragement of content sharing may be done by implementing automatically the share of completed downloads by default. This means that when the download is completed, the file just downloaded should be shared automatically by the downloading users. Download Failures may be solved by having an incomplete download directory to prevent the sharing of partial files.

There are two types of freeloaders: the users who only download files for themselves without ever providing files for download to others, and the users who provide low-quality contents to others for download. Blocking freeloaders can be solved by blocking web-based downloads; blocking queries from web-based indexing tools, or using reward sharing and punish freeloaders method (this may require slight changes in the protocol).

Reducing unnecessary traffic attempts to reduce the number of ping and pong messages. One approach to accomplish this is not to periodically ping the network, but may be set the rule that the ping message can be sent only when the directly connected peers are left only two or lower. Creating and maintaining the healthy network can be done by dropping a connection that is going to get clogged (this may remove the slower modem users from the center or fast areas of the networks) and using the prioritization scheme to discard the message if necessary. The rule of dropping a connection may be of the following form: dropping a connection if it has been unable to receive the message for about 10 seconds, and dropping a connection if it has sent no message for about 10 seconds and does not respond to a ping with TTL = 1. The prioritization scheme is used to selectively drop the most useless message first if necessary. The priority of messages ordered from high to low is: push > queryHit > query > pong > ping.

Another approach to improve performance is through the use of Super-peers (this is used in newer P2P architectures like FastTrack and OpenFT) and caching of the results. One cause of the performance degradation comes from the assumption that every servent has the same resource capacity, but in fact, many servents have weak connections (low-speed modem), and participate in the network only for a short time interval. This makes these servents unsuitable for taking the active part of the protocol. Super-peers can be used as proxies for less powerful servents. Super-peers, which typically have high bandwidth, act as local search hubs on behalf of the less powerful peers connecting to them and they always stay on-line permanently. Super-peers also help reduce network traffic by caching query routing information. Actually, Gnutella v0.6 protocol already proposed the use of Super-peers. The use of Super-peers will make the Gnutella network topology be the mix of centralized and decentralized network as illustrated in the figure below.

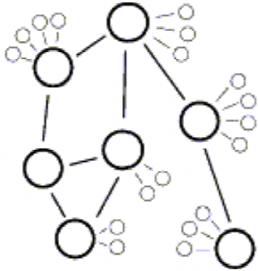

Figure 23: Centralized and Decentralized network topology (Nelson, 2001)

To reduce the time used to find the matching of query, each servent can cache a queryHit sent back passing him associated with the query sent to him before. In case when there is the same query associated with that queryHit sent to him again, he can send back the queryHit corresponding to that query immediately. This method will perform best when repeated queries are common. We can make this effect by arranging the servents interested in the same contents staying closed together in the network. Note that pong messages can also be cached in the same way so that the next time the ping messages come, a servent can return the cached pong messages immediately eliminating the need to forward this ping message further. The idea is that each servent remembers all pong messages that it has received. When his neighbor sends a ping message with TTL of n to him, he will reply with his pong message and all its cached pongs with hop count less than n. This will finally result in each ping message propagated to only the directly connected neighbors, and servents will not need to forward the same pong messages repeatedly.

### 4.3.3 Reliability
The third issue, reliability, has been a hot discussion recently. Reliability focuses on the security issues because malicious users can make the network from functioning properly implying that the reliability is compromised. Since the Gnutella network is in distributed form, the contents are spread variously in many nodes, the attack of specific machine as in client-server model cannot make Gnutella network down. However, the introduction of Super-peers may help attackers to possibly perform this kind of attack by targeting at the Super-peers. Even though this will not make the entire Gnutella network down, some parts of the network can suffer significantly. Apart from that, there are three kinds of attacks that can be done easily in the loosely structured, highly dynamic P2P network like Gnutella: Flooding, Malicious or Fake Contents, and Hijacking queries. This will be discussed based on (Dimitri, 2002). In fact, Gnutella network opens many security risks since the design focus is primarily on functionality and efficiency.

Flooding, which can lead to Denial of Service, can caused by malicious users' producing a large number of query messages or supplying erroneous replies to all queries. Note that flooding of pong or queryHit messages is useless since they will be discarded if there are no matching ping or query messages sent over the same network previously. Flooding of ping messages is also unfruitful because of the caching of pong messages discussed previously. The author of (Dimitri, 2002) said that flooding problem can not be prevented but can minimize by using load balancing.

Malicious or Fake contents come in many forms, such as the file downloaded may contain virus, content alteration, or the content of the file is not what is expected or advertised. What is worse is that the virus files, if they exist in the network, tend to propagate quickly. Therefore, it is essential to have a way to authenticate the content of the file before receiving it. The author of (Dimitri, 2002) suggested the Fourier Transforms technique to determine the content in the file, that are not bit-wise identical, to see whether it is what is expected.

Hijacking queries can be done easily in the Gnutella network. This stems from the trust of intermediate third parties in the network. The use of Super-peers also makes it easy to hijack the queries since each Super-peer can see a large portion of queries. Point-to-point encryption, Distributed Hash Tables, and Selection mechanism are all the techniques proposed by (Dimitri, 2002) to help resist flooding and query interception attacks.

According to (Matei), the current Gnutella network follows a multi-modal distribution form, a combination of a power-law and quasi-constant distribution. This topology keeps the network reliable almost as a power-law distribution topology; the network can tolerate random node failures.

Unfortunately, Gnutella network topology can be acquired easily by using the crawler and analyzing the data gathered. This would permit highly efficient denial of service attack by educated attackers.

### 4.3.4 Anonymity

The last issue, anonymity, can be achieved naturally in the Gnutella network. The requested servents will not know who the requesters are since the requesters can be anyone along the query paths. In the same way, the requesting servents do not know who the requested servents are. However, this anonymity makes possible many security threats to occur easily. Unfortunately, anonymity can be invaded by using the Descriptor ID in the Descriptor header to trace back Gnutella messages.

## 5. FastTrack

One of the newer and more inovative peer-to-peer architectures would be the FastTrack network. It came as a solution to the problems that both Napster and Gnutella was facing. The FastTrack network, is by nature a Hybrid Architecture which, as mentioned in the earlier sections, is a cross between two or more basic network topologies. For FastTrack, it is namely the cross of the centralized and decentralized topologies. (Yang, 2002).

## 5.1 The FastTrack Protocol

The FastTrack protocol is a proprietary architecture, where rights to use the network has to be obtained through a company called Sherman Networks (Marcus, 2003). Therefore, very little is known of the actual protocol used. Many attempts have been made to reverse engineer the FastTrack protocol. The most well known to date would be the giFT project, as they were the closest to finally cracking the protocol. FastTrack reacted by changing it's encryption, to the point where it was virtually impossible to reverse engineer (Marcus, 2003). The work done by the giFT project however, was sufficient to give a rough outline of how the FastTrack protocol actually works, even if the information may now very well be outdated. The following section contains a description of the architecture used by the all FastTrack clients.

This technology uses two tiers of control in its network. The first tier is made up of clusters of ordinary nodes that log onto Super Nodes (ordinary machines with high speed connection). As discussed previously, this sort of connection mimics the centralized topology. The second tier consists of only Super Nodes that are connected to one another in a decentralized fashion.

The number of peers that can be designated as Super Nodes can vary from tens to several thousand. This is because these Super Nodes themselves are just ordinary nodes that can and will join or leave the network as they please. Therefore, the network is dynamic and always changing. In order to ensure the constant availability of the network, there exist a need for a dedicated peer (or several of these peers) that will monitor and keep track of the network. Such a peer is called a bootstrapping node (Kurose, 2003) and it should always be available online. When a FastTrack client, for example Kazaa is executed on a peer, it will first contact the bootstrapping node. The bootstrapping node will then determine if that particular peer qualifies to be a Super Node. If it does, then it will be provided with some (if not all) IP addresses of other Super Nodes. If it only qualifies to be an ordinary peer, then the bootstrapping node will respond by providing the IP address of one of the Super Nodes (Hari, 2002).

Certain FastTrack clients like Kazaa Lite uses a method known as the 'Reputation System', where the reputation of a certain user is reflected by their participation level (a number between 0 and 1000) in the network. The longer the user stays connected to the network, the higher their participation level will be, which in turn means that they will be more favoured in queuing policies and hence should receive better service. This is mainly to encourage users to share files and thus effectively reducing the number of 'passenger clients' on the network.

Resource discovery is accomplished through the act of broadcasting between Super Nodes. When a node from the second tier makes a query, it is first directed to its own Super Node, who will in turn broadcast that same query out to all other Super Nodes that it is currently connected to. This is done repeatedly until the TTL (Time To Live) of that query reaches zero (Ashish). So, if for example, the TTL of a query is set to 7 and the average amount of nodes per Super Node is 10, a FastTrack client is able to search 11 times more nodes on a FastTrack network as compared to Gnutella (Duncan, 2001). This gives FastTrack clients a much greater coverage and hence, better search results. There is one drawback to such a broadcasting method, and that is the daunting amount of data that needs to be transferred from Super Node to Super Node. This is the very same problem that has been plaguing the Gnutella network. This, however, is not a serious problem for Kazaa as opposed to Gnutella, for the Super Nodes are nodes that are guaranteed to have fast connections.

Each of the Super Nodes that received the query will then perform a search through its indexed database that contains information of all the files shared by its connected nodes. Once a match is found, a reply will be sent back following the same path the search query was propagated through until it reaches back to the original node that issued the query. This method of routing replies is similar to Gnutella, and hence runs the risk of facing the same problem of losing replies as it is routed back through the network. This is due to the fact that the Gnutella network backbone, as mentioned previously, is made up of peers that connect and disconnect from the network very sporadically. This would mean that reply packets that are being routed back may be lost as the path it took is no longer there because one or more of the nodes making up the link disconnected itself. The afore mentioned problem however, is not as serious for Kazaa users as the FastTrack network backbone is made up of peers that have high speed connections (Super Nodes) and hence the return paths can be considered more reliable.

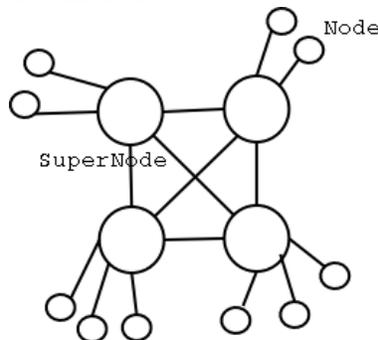

Figure 24: Illustration of the FastTrack Topology

## 6. OpenFT

As mentioned in the previous section, the FastTrack network protocol is proprietary hence not much about its protocol is known. This very fact has become the source of motivation for various individuals and groups to try and break the FastTrack protocol. The most well known of all, would be the giFT project. According to (Marcus, 2003), the initial meaning of the abbreviation giFT was Generic Interface to FastTrack or giFT isn't FastTrack. The giFT project came very close to actually breaking the FastTrack protocol. However, FastTrack reacted by changing their encryption, making it impossible to reverse engineer.

Since the FastTrack protocol was no longer feasible to crack, the aim of the giFT project was changed to develop a system which can connect many different heterogeneous networks and still be able to share files between them. Therefore, the meaning its abbreviation was also changed to giFT Internet File Transfer (Marcus, 2003).
The project lead to the development of a new and improved network protocol that was very much like FastTrack; it was called OpenFT.

## *6.1 The OpenFT Architecture*

Like FastTrack, the OpenFT protocol also classifies the nodes in its network into different roles, but instead of a two-tier control architecture, OpenFT has added an extra tier making it a three-tier control architecture. The classification of nodes is done based on the speed of its network connection, its processing power, its memory consumption and also its availability (SourceForge, 2002).

The first tier would be made up of clusters of ordinary machines which we refer to as User Nodes. These nodes themselves, maintain connections to a large set of Search Nodes (ordinary machines with high speed connection). The user nodes will then update a subset of the search nodes that it is connected to with information regarding files that are being shared (SourceForge, 2002).

The second tier is made up of machines that are referred to as Search Nodes. These nodes are the actual servers in the OpenFT network. These servers have the responsibility to maintain indices of files that are shared by all the user nodes under them. On default, a search node can manage information about files stored at 500 user nodes (SourceForge, 2002).

The third tier is made up of a group that it much smaller, because the requirements to qualify for this group are much more stringent. One has to be a very reliable host that has to be up and available most of the time. These nodes are referred to as Index Nodes as their main purpose is to maintain indices of existing Search Nodes. They also perform tasks like collecting statistics and monitoring network structure. Basically, this group can be seen as the administrator group that ensures all other participating nodes are working fine and are up to expectation (SourceForge, 2002).

It should be noted that the second and third tier of control can actually be performed by the same node. In other words, a node can function both as a Search Node and as an Index Node at the same time.

By default, each User Node has to select 3 Search Nodes to maintain their shared file information. If accepted, the selected Search Node will be a Parent to that particular User Node, who will now have to send the list of its shared files to it (Marcus, 2003).

Having said that, all nodes will also have to maintain a list of available Search Nodes. When a query is sent, it will be sent to all nodes that are found in this list (SourceForge, 2002).

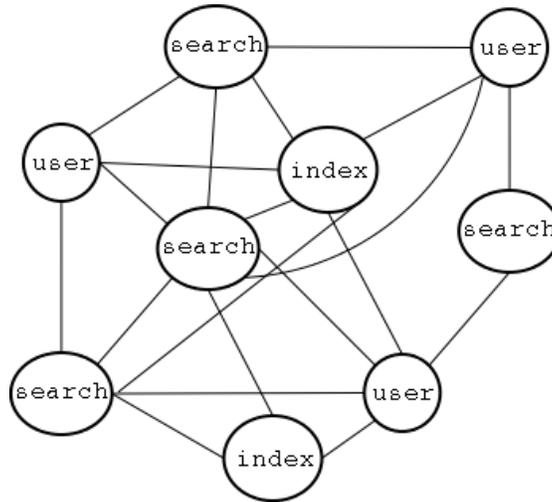
Figure 25: Illustration of the OpenFT Topology

## 6.2 The OpenFT Protocol

The OpenFT network protocol uses a simple packet structure to make it easy to parse. There are all together 28 packet types and all of them have the following packet header.

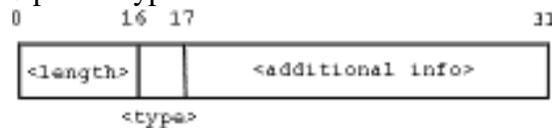
Figure 26: Illustration of the OpenFT Header

Length describes the length of the packet excluding the header. The current implementation uses bits 16-31 to store both flags and packet type information (Marcus, 2003).

As mentioned in the previous section, each node is assigned a particular role in the network. At the moment of writing, there is still yet a distributed algorithm that can be implemented to perform the role assignment task. For now, the users themselves choose the role it wants to play in the network.

The table below gives a summary of the packets used by the OpenFT network protocol (Marcus, 2003; Josh).

| Packet | Description |
| --- | --- |
| version | Used to indicate protocol version and check to see if it is outdated. |
| nodeinfo | Used to communicate node information such as IP address, port numbers and OpenFT node category. |
| nodelist | Used to list down other nodes. |
| nodecap | Used to indicate node capabilities. |
| ping | Used to keep connection alive. |
| session | Used to establish session. |
| child | Used to request to be the child of a particular Search Node. |
| addshare | Used to add information about the files that are shared in a User Node to the registry of a Search Node. |
| remshare | Used to remove information about the files that are shared in a User Node to the registry of a Search Node. |
| modshare | Used to modify information about the files that are shared in a User Node to the registry of a Search Node. |
| stats | Used to request network statistics from Index Nodes. |
| search | Used to query for files. |
| browse | Used to browse the files shared by a particular node. |
| push | Used to perform HTTP PUSH (like the one mention in Gnutella) through firewalls. |

Table 3: Summary of packets used in the OpenFT network protocol

All the packets mentioned can basically be categorized into several categories so that their usage and function can be seen more clearly. With reference made to (Marcus, 2003), they are:-
- Session Establishment: version, nodeinfo, nodelist, nodecap, session
- Connection Maintenance: ping
- File Sharing Management: child, addshare, remshare, modshare
- Searching for files: search, browse
- Misc: stats, push

# 7. COMPARISONS

| Purpose | Napster | MP3 file sharing. |
|---|---|---|
| | Gnutella | File sharing of all types. |
| | FastTrack | File sharing of all types. |
| | OpenFT | File sharing of all types. |
| Architecture | Napster | Peers connected to a centralized server |
| | Gnutella | Flat/Ad-hoc network of peer servents (Pure P2P) |
| | FastTrack | Decentralized two-level hierarchical network of group-leader peers and ordinary peers |
| | OpenFT | Decentralized three-level hierarchical network of search peers, index peers and ordinary peers |
| Lookup Method | Napster | Using central directory |
| | Gnutella | Query flooding: Broadcast queries to peers and make a direct connection when download. |
| | FastTrack | Using group-leader peers |
| | OpenFT | Using the search peers |
| Decentralization | Napster | The system is highly centralized. Peers are connected directly to the central index server. |
| | Gnutella | The system is highly centralized. The topology is flat and each peer is truly equal. |
| | FastTrack | The system is decentralized in the sense that there is no explicit central server. However, in each group, the ordinary peers are still connected to their group-leader peer in a centralized manner. |
| | OpenFT | The network structure is very much like FastTrack, only OpenFT is slightly more optimized for performance. |
| Scalability | Napster | The system is not scalable because having every one connected to the directory server is bottleneck prone. |
| | Gnutella | Theoretically, the system can easily expand indefinitely but in practice may be limited by the underlying bandwidth. Traffic jams of a large number of messages transmitted over the network can cause the network to stop functioning properly. |

|  | FastTrack | The scalability of the system is medium. Although the architecture is fairly decentralized, it is not entirely serverless. A bottleneck may occur locally within a group. |
|  | OpenFT | The scalability of the system is just like FastTrack. It is easily scalable, but bottlenecks may occur within a group since it is not entirely serverless. |
| Anonymity | Napster | Hardly any anonymity since all users have to sign up onto the central server before they can connect to the network. Users are almost always anonymous to each other though, since the user directory is never queried directly. |
|  | Gnutella | Anonymity is preserved naturally by the protocol itself since the messages may come from any peers in the local paths. However, the Descriptor ID in the Descriptor header may be used to trace back the messages, intermediate. Nosy nodes can record queries, responses. |
|  | FastTrack | Level of anonymity is slightly better than Napster but less than Gnutella. Users are not fully anonymous in the FastTrack network. |
|  | OpenFT | Level of anonymity is similar to FastTrack. |
| Security | Napster | Moderate, since Napster is managed by a centralized server. This allows Napster to exert much more control which makes it much harder for clients to fake IP addresses, port numbers, etc. And since Napster only allows the sharing of MP3 files, this makes security threats that are related to fake content, viruses, etc. are not too prone and more traceable as compared to the other three protocols. |
|  | Gnutella | Low, since the protocol design primarily focuses on the functionality and efficiency. Prone to several security threats: flooding, malicious or fake content, virus spread, hijacking of queries, denial of service attacks. Security threats mainly focus on services rather than host. In the case of one user staying behind the firewall, download can be achieve by requesters sending a push message to ask the firewalled user to initiate a connection. In the case of two users both staying behind firewall, a connection cannot be created. |
|  | FastTrack | Low, but is still better than Gnutella, since it maintains a hybrid architecture. Security threats such as flooding, malicious or fake content, |
| | | viruses, etc. can be reduced as all the Super Nodes actually function as centralized servers to nodes that are under their domain. Another threat is the integration of spyware into the popular FastTrack client, Kazaa. These spywares monitor the activities of users in the background. Any information from surfing habits to favourite web sites can be gathered and sent back to the server, who will then use this information for targeted marketing, etc. |
| | OpenFT | Very similar to threats faced by FastTrack. One of the main differences is that it doesn't have spywares integrated into its clients. |
| Self-organization | Napster | A highly self-organized system is not necessary. Organization of nodes/resources are handled by the central servers. |
| | Gnutella | The system is highly self-organized. It adapts gracefully with the dynamic nature of the Internet through the use of ping messages to periodically find the available peers connected to the network. But this comes with a drawback, which is the staggering amount of data transfer involved. |
| | FastTrack | The system is very much self-organized as well. It adapts to the dynamic nature of the Internet, just like Gnutella. The slight difference in its protocol allows FastTrack to achieve a highly self-organized system with much less data transfer as compared to Gnutella. |
| | OpenFT | Self-organization in OpenFT is similar to FastTrack. |
| Lookup Completeness | Napster | The lookup method is complete, because search is done by the central server which has the complete index directory. |
| | Gnutella | The lookup method is incomplete, because a query may not reach all the servents. I has a much greater coverage compared to Napster, but it takes a much longer time to perform a lookup. |
| | FastTrack | The lookup method in FastTrack is incomplete as well, but it is able to search many times more nodes as compared to Gnutella for the same time span. |
| | OpenFT | The lookup method on OpenFT is similar to FastTrack. There maybe some minor adjustments to improve performance, but on the baseline, they are both the same. |

| Fault Resilience | Napster | The malfunction of the central server can cause a system wide malfunction. |
|---|---|---|
| | Gnutella | The malfunction of some nodes would not cause the system to stop functioning as long as there are enough nodes connected to the network at a given time. But this will definitely degrade performance. |
| | FastTrack | The malfunction of an ordinary peer would not hurt the system. The malfunction of a group-leader peer is taken care by re-assigning all ordinary peers connected to other group-leader peers. |
| | OpenFT | Fault resilience for OpenFT is similar to FastTrack, where the malfunction of a user node would not hurt the system. If any of the index nodes or search nodes fails, all that needs to be done is just re-assign all user nodes affected to other index or search nodes. |

## 8. Summary

It should be relatively clear by now that P2P technology is still in its infant stage of its development. There is still great potential for growth and improvements that can be done. From this chapter alone, we can see how P2P has evolved from a more centralized architecture like Napster into a fully distributed architecture like Gnutella; only to evolve again into a hybrid architecture like FastTrack and OpenFT. This evolution in technology is spurred mainly by the need to achieve better efficiency and speed in content sharing as well as for the need to survive law suits against these architectures. Many more innovative architectures will surface as the race toward network efficiency and survival continues.